\documentclass[twocolumn,epjc3]{svjour3}  
\UseRawInputEncoding
\smartqed   
\RequirePackage{xcolor}
\RequirePackage[colorlinks,linkcolor=blue,citecolor=blue,urlcolor=blue]{hyperref} 
\RequirePackage{newtxtext,newtxmath}
\RequirePackage{doi}
\RequirePackage{graphicx}
\RequirePackage{multirow}
\RequirePackage{array}
\RequirePackage{mathtools}
\RequirePackage{amsmath}
\RequirePackage{amsfonts}
\RequirePackage{tikz}
\usetikzlibrary{positioning}
\renewcommand{\arraystretch}{1.5}
\RequirePackage{bm}
\allowdisplaybreaks
\usepackage{rotating}
\usepackage{lscape}
\usepackage{array}
\usepackage{lipsum}

\hypersetup{
    colorlinks=true,
    linkcolor=red,
    citecolor=blue,
}

\RequirePackage[numbers,sort&compress]{natbib}

\newcolumntype{M}[1]{>{\hbox to #1\bgroup\hss$}l<{$\egroup}}

\makeatletter
\newcommand\@brcolwidth{0.67em}

\def\@brarray[#1]{\array{r*\c@MaxMatrixCols {M{#1}}}}
\makeatother

\journalname{Eur. Phys. J. C}
\begin{document}

\title{Detection of gamma-ray burst Amati relation based on Hubble data set and Pantheon+ samples
}

\author{Yufen Han\thanksref{addr1}\and
      Jiaze Gao \thanksref{addr1}\and  
      Gang Liu \thanksref{addr1}\and
      Lixin Xu\thanksref{e2,addr1} 
}

\thankstext{e2}{e-mail: lxxu@dlut.edu.cn (corresponding author)}

\institute{Institute of Theoretical Physics, School of Physics, Dalian University
of Technology, Dalian 116024, People's Republic of China \label{addr1}
}

\date{Received: date / Accepted: date}

\maketitle

\begin{abstract}

 gamma-ray bursts as standard candles for cosmological parameter constraints rely on their empirical 
  luminosity relations and low-redshift calibration. In this paper, we examine the Amati relation and its potential 
  corrections based on the A118 sample of higher-quality gamma-ray bursts, using both Hubble data set and 
  Pantheon+ samples as 
  calibration samples in the redshift range of $z<1.965$. In calibrating gamma-ray bursts using these two datasets, 
  we employ Gaussian processes to obtain corresponding Hubble diagrams to avoid the
  dependence on cosmological models in the calibration process. We first divided the low-redshift sample of GRBs into 
  two bins and examined the Amati relation and its potential modifications. We found that under both calibrations, 
  the Amati relation did not show evidence of redshift evolution (68$\%$ confidence level). For the other two Amati 
  relations that include redshift evolution terms, the central values of the redshift evolution coefficients deviated 
  from 0, but due to the limitations of the sample size and the increase in the number of parameters, most of the redshift 
  evolution coefficients were not able to be excluded from 0 at the 1$\sigma$ level. Therefore, to assess their 
  situation across the entire redshift range, we 
  employed MCMC to globally fit three types of Amati relations. By computing AIC and BIC, we found that for the GRB 
  A118 sample, the standard Amati relation remains the most fitting empirical luminosity formula, and no potential 
  redshift evolution trend was observed for two different low-redshift calibrating sources.
\end{abstract}

\section{Introduction}
\label{intro}

In 1998, two independent research groups measured the relation between the distance and redshift of type Ia 
supernovae to accurately determine the matter density within the range of $0<z<0.7$, which showed that the universe 
was in an accelerating expansion phase~\cite{Alexei1998OBSERVATIONALEF,perlmutter1999measurements}. Since then, 
other observations such as cosmic microwave background radiation (CMB)~\cite{spergel2003first,collaboration2014planck,
ade2016planck,akrami2020planck} and baryon acoustic oscillations (BAO)~\cite{eisenstein2005detection,alam2017clustering} 
have confirmed that the universe is in an accelerating expansion. The accelerated expansion of the universe implies that the current 
cosmic evolution is predominantly governed by a substance with negative pressure, known as dark energy. 
However, our understanding of the dark energy component dominating the current universe evolution remains 
incomplete to date, necessitating the exploration of cosmological probes at higher redshifts.

The mechanism of standard type Ia supernovae (SNe Ia) limits their maximum luminosity, with observed redshifts 
typically around $z$ $\sim$ 2.3~\cite{scolnic2018complete}, which can be used as a probe for studying cosmology 
\cite{Scolnic_2018,Gao_2020,Brout2022,gao2023measurementhubbleconstantusing,dainotti2024reduceduncertainties43hubble}. 
Clearly, to extend cosmological studies to higher 
redshifts, gamma-ray bursts (GRBs) can serve as complementary cosmic probes. 
The maximum redshift of gamma-ray bursts observed so far can reach approximately $z \sim 9.4$~\cite{cucchiara2011photometric}, 
with luminosities ranging from $10^{47}$ erg $\mathrm{s^{-1}}$ to 
$10^{54}$ erg $\mathrm{s^{-1}}$~\cite{zhang2011open}.

When utilizing gamma-ray bursts as cosmological supplementary probes, this is based on some empirical luminosity 
correlations for their standardization
~\cite{fenimore1999gamma,norris2000connection,amati2002intrinsic,FRAIL2004255,ghirlanda2004collimation,
yonetoku2004gamma,liang2005model,firmani2006discovery,dainotti2008time,liang2008cosmology,
tsutsui2009cosmological,xu2012new,liu2022improved}. Additionally, ~\citet{amati2019addressing} utilized 
Hubble observational data to approximate the Hubble function through a linear combination of Bernstein basis 
polynomials, obtaining Bézier parameter curves. They employed Observable Hubble Data (OHD) to construct luminosity 
distance in a model-independent manner. ~\citet{Ghirlanda2010SpectralEO} employed gamma-ray bursts 
with a narrow range of redshifts to calibrate the $E_p - E_{\gamma}$ (gamma-ray energy corrected for beaming) 
relation. ~\citet{xu2012cosmological} utilized a model-independent method to calibrate GRBs, primarily employing 
internally calibrated gamma-ray bursts for measurements.

However, during the calibration of gamma-ray bursts, ensuring their standardizability is crucial, 
as noted in Ref. \cite{osti_1671149}. When employing empirical luminosity relations for GRB 
calibration, some analyses rely on cosmological models, thereby biasing constraints on cosmological 
parameters towards assumed cosmological models. 
\citet{osti_1671149}
analyzed GRB data across six different cosmological models and revealed that the Amati relation remains 
independent of cosmological models. This indicates that this sample of GRBs can be calibrated using the 
Amati relation, thereby enabling its application in cosmological studies.
  
When using calibrated gamma-ray bursts for cosmological studies, overcoming the so-called 
circularity problem is essential. One approach is simultaneous fitting, where GRBs are 
jointly fitted with other cosmological phenomena. This approach entails the simultaneous 
fitting of relevant parameters within GRBs and parameters of the cosmological model \cite{ghirlanda2004gamma,
2007Overcoming,amati2008measuring,postnikov2014nonparametric,osti_1671149,Cao_2020,khadka2021gamma,
Cao_2021,cao2022gamma,cao2022standardizing,Cao_2022,dainotti2023gamma,Cao_2023}. 
Additionally, the distance ladder approach can be employed. When employing the distance ladder approach to calibrate GRBs, it is crucial to select an appropriate 
standard probe. This assumes that the chosen probe maintains a standardized absolute value consistently from the 
proximal to the distal ends of the cosmic distance ladder. In accordance with~\citet{dainotti2008time,
dainotti2010discovery,dainotti2011study}, it is considered that gamma-ray bursts exhibit a luminosity distance-redshift 
relation consistent with that of type Ia supernovae at low redshifts, and this relation is subsequently extended to the 
high-redshift regime of GRBs.

Calibrating gamma-ray bursts (GRBs) using the distance ladder concept, the correlation coefficient of GRBs low 
redshift empirical luminosity relation is obtained. If it is directly extended to high redshift, it is ignored 
that the correlation coefficient of empirical luminosity relation may evolve with redshift. For example, 
~\citet{10.1093/mnras/stv2471} found possible redshift evolution in several correlations (including 
$E_p - E_{\gamma,iso}$ correlation), but no significant redshift evolution was found in the $E_p - E_{\gamma} $ 
relation. Furthermore, Xu et al.~\cite{Xu_2021} expanded the Dainotti relation by incorporating a redshift 
evolution term. By assessing the coefficient of this term, they investigated whether this relation evolves with 
redshift. Then, following the approach outlined in Ref. \cite{Dainotti_2020}, they employed the Efron-Petrosian 
method~\cite{Efron1992AST} to remove potential evolution in this parameter.

When calibrating the gamma-ray burst Amati relation with a low-redshift probe, taking into account that the obtained 
correlation coefficients at low redshift may evolve with redshift,~\citet{wang2017evolutions} and~\citet{demianski2021prospects} 
extended the Amati relation differently by adding a redshift evolution term to 
describe the redshift evolution of the photometric empirical relation by calibrating the Amati relation with the 
GRBs. The obtained correlation coefficients at low redshifts were expanded to high redshifts to obtain the 
distance modulus at high redshifts of the gamma-ray bursts, and the calibrated gamma-ray bursts were used to 
constrain the cosmological parameters. In this paper, we will also calibrate gamma-ray bursts based on the concept 
of the distance ladder, assess the redshift evolution of the Amati relation, and explore the Amati relation and its 
potential modifications across different redshift bins. Additionally, we will discuss and analyze which of these 
three empirical luminosity relations better fits the data points.

In this paper, we reconstructed the relation between redshift $z$ and distance modulus $\mu$ using Gaussian 
processes based on the Hubble dataset and the SNe Ia Pantheon+ samples, obtaining the values and errors of the 
distance modulus for GRBs in the redshift range from 0.0331 to 1.965. We performed binning on low-redshift data 
of gamma-ray bursts and obtained the correlation coefficients of the Amati relation and its extended version for 
different bins using MCMC sampling. We then assessed the fitting results of the three relations to the data. 
We further performed a global fit of the Amati relation and its extended version using MCMC sampling. By 
calculating AIC and BIC, we selected the empirical luminosity relation that provides a better fit to the data. 
We then extended the correlation coefficients to higher redshifts, obtaining the distance modulus corresponding 
to the redshifts of high-redshift gamma-ray bursts. Additionally, we extended the Hubble diagram to higher redshifts. 

The structure of this paper is as follows. In Section~\ref{sec:2}, we reconstruct the relation between 
redshift and distance modulus of two kinds of data, namely Hubble dataset $H(z)$ and Pantheon+ samples, respectively by Gaussian 
process, and obtain the distance modulus of low redshift corresponding to GRBs. In Section~\ref{sec:3}, we 
calibrate the Amati relation of gamma-ray bursts and its potential modifications. We calculate, analyze, and discuss 
these three relations separately. By selecting the appropriate relation and applying its corresponding coefficients, 
we obtain the distance modulus for high-redshift gamma-ray bursts. This allows us to extend the Hubble diagram to 
higher redshifts. Section~\ref{sec:4} summarizes and discusses the obtained results.

\section{Data selection and reconstruction of the relation between low redshift and distance modulus}
\label{sec:2}
\subsection{GRBs}

In this paper, we selected the A118 sample from Ref. \cite{khadka2021gamma} for our gamma-ray burst dataset 
because it has been demonstrated therein to be standardized, possessing higher quality, and more 
suitable for cosmological research \cite{khadka2021gamma,cao2024testingstandardizabilityofderiving}.  
This sample comprises 118 long GRB data points with redshifts 
ranging from 0.3399 to 8.2000. Notably, it has been noted that there 
is an error in the $E_p$ value for the GRB081121 data point \cite{liu2022improved,Cao_2022}.
Therefore, in this paper, we utilize 
the corrected value as provided in Ref. \cite{Wang_2015}.

Due to the use of low-redshift calibrated astrophysical objects, with a maximum observed redshift of 1.965, we 
can reconstruct the corresponding distance modulus for each redshift based on cosmological principles. Consequently, 
we can also derive the distance modulus values for gamma-ray burst data points with redshifts lower than 1.965, 
comprising a total of 48 data points. At the same time, in order to determine whether the empirical luminosity 
relation of gamma-ray bursts evolves with redshift at low redshifts and to explore the differences among these 
relations at different redshift ranges, we divided these 48 GRB samples into two parts. The low-redshift subset 
consists of 24 data points, with a redshift range of $0.3399<z<1.46$, while the high-redshift subset comprises 
24 data points, with a redshift range of $1.48<z<1.95$.

\begin{figure*}[t]
  \centering
  \includegraphics[width=1.0\textwidth]{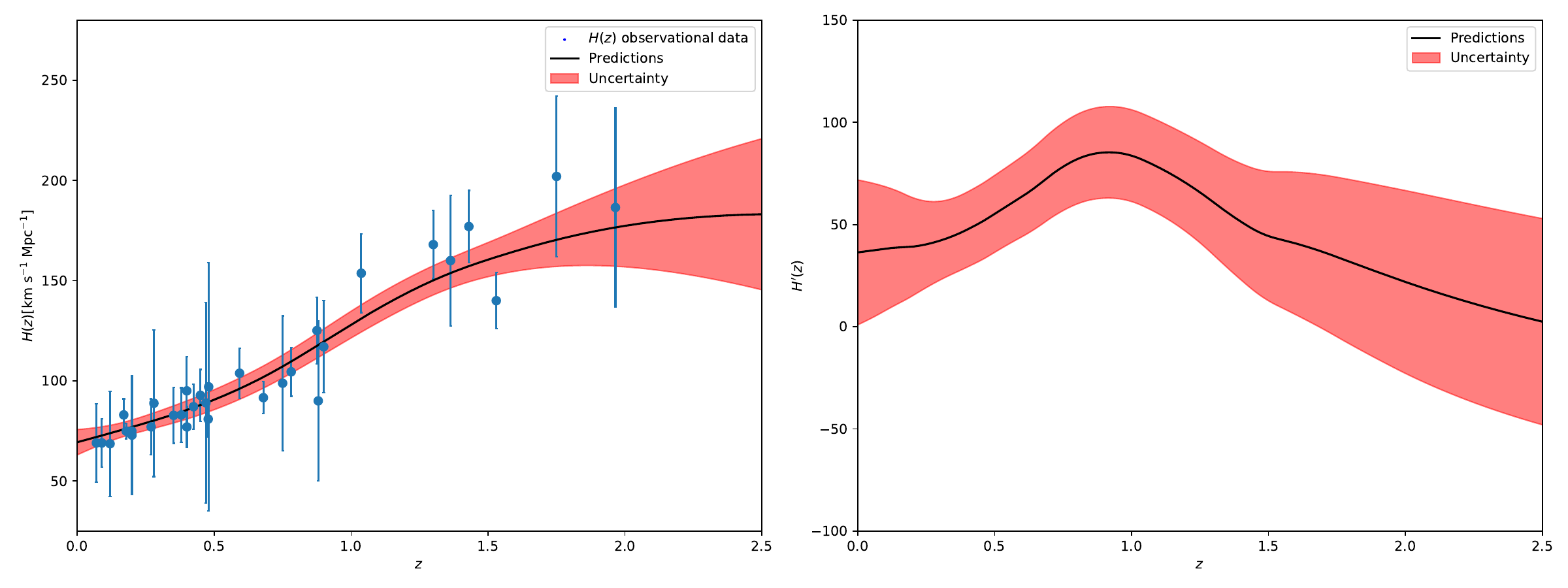}
  \caption{{The reconstruction of the $z-H(z)$ plot using GaPP. In the plot, the blue dots represent the data points 
  of the Hubble parameter, $H(z)$, along with the associated 1$\sigma$ error bars. The black line represents the 
  reconstructed Hubble data using GaPP, while the shaded red region denotes the 1$\sigma$ error of the 
  reconstructed data.}}
  \label{fig:1}
\end{figure*}
\subsection{$H(z)$}
The Hubble parameter characterizes the relative expansion rate of the universe and encapsulates valuable 
information regarding the cosmic expansion history, rendering it a crucial cosmological observable. 
In this paper, we utilized the updated Hubble parameter data from Table I in reference \cite{Cao_2023} 
to calibrate gamma-ray bursts. The dataset comprises 32 data points within a redshift range of
$0.07 < z < 1.965$, providing values for $z$, $H(z)$, and $\sigma_{H(z)}$. Notably, 15 of these 
measurements are correlated. In subsequent analyses, we employed the complete covariance matrix.

In a spatially flat universe, the formula for luminosity distance is as follows:
\begin{equation}
    d_L = {c(1+z)\int_{0}^{z} {1\over H(z')} \,dz'}. 
    \label{eq:1} 
\end{equation}
Here, $c$ represents the speed of light, and $z$ denotes redshift, which can be 
calculated using Hubble data to obtain the luminosity distance corresponding 
to each redshift value.

\subsection{SNe Ia}

Type Ia supernovae stand as indispensable celestial phenomena within the domain of cosmological investigations, 
wielding significant influence in our quest to unravel the mysteries of the universe. In this paper, we use the 
Pantheon+ samples\footnote{\url{https://github.com/PantheonPlusSH0ES/DataRelease}.} of SNe Ia, which consists of 
1701 light curves of 1550 spectroscopically confirmed Ia supernovae 
from 18 different sky surveys, with redshifts in the range of $0.00122<z<2.26137$~\cite{osti_1893974,
Brout_2022}.

In a spatially flat universe, the distance modulus of a supernova observation from an observational point of view is:
\begin{equation}
  \mu_{SN}= m_B - M ,
  \label{eq:2}
\end{equation}
where  $m_B$ is the apparent magnitude, $M$ is the absolute magnitude. In this paper, we use a Gaussian prior from 
the Cepheid calibration distance ladder method.  As discussed by Efstathiou 
~\cite{10.1093/mnras/stab1588}, the SH0ES prior should be applied to $M$ rather than $H_0$ when combining SH0ES data with 
other astrophysical data in order to constrain the late-time physics.  We selected the result $M = -19.253\pm0.027$ 
from \citet{Riess_2022} calibrating SNe Ia via Cepheid. Since $M$ is a posterior derived from the Cepheid 
calibration of the Pantheon+ samples with redshift $z<0.01$, we selected 1590 data points with redshift $z>0.01$ for use.

\subsection{Reconstructing the relation between low redshift and distance modulus via the Gaussian process}

In recent years, Gaussian processes have found extensive application in cosmology~\cite{shafieloo2012gaussian}.
Gaussian processes are an extension of Gaussian distributions, where Gaussian distributions describe the random 
distribution of variables, whereas Gaussian processes aim to construct the distribution function of random 
variables~\cite{seikel2012reconstruction}. In Gaussian process, the joint distribution of any finite random 
variables is a multivariate Gaussian distribution, and the covariance between different random variables is 
determined by a kernel function. Specifically, a Gaussian process can be fully described by a mean function and a 
covariance function. Given any set of input data points, the Gaussian process can generate the corresponding output 
value, and the prediction of any input data point is accompanied by an estimate of uncertainty.
Thus, without assuming a specific parametric form, Gaussian processes can reconstruct 
the function from data points via a point-to-point Gaussian distribution \cite{seikel2012reconstruction}.

In this paper, we use GaPP\footnote{\url{https://github.com/carlosandrepaes/GaPP}.} \cite{seikel2012reconstruction} to implement Gaussian processes.
GaPP is an open-source Python tool that provides methods for Gaussian process regression, allowing 
reconstruction of functions and their first, second, and third-order derivatives from observed data. 
Next, we proceed with a brief introduction to Gaussian processes. Observational 
data can be described using Gaussian processes, for a Gaussian noise $\epsilon_i$ with 
variance $\sigma_i^2$, the actual observation $y_i=f(x_i)+\epsilon_i$, whose Gaussian 
process can be written as:
\begin{equation}
  \boldsymbol{y}\sim \mathcal{N} (\boldsymbol{\mu} ,K(\boldsymbol{X, X})+C).
  \label{eq:23}
\end{equation}
Here, $\mathcal{N}$ represents the Gaussian process, the covariance matrix $K(\boldsymbol{X, X})$ is 
derived from the kernel function $k(x_i, x_j)$, $\boldsymbol{\mu}$ denotes the mean of the 
reconstructed function, and $C$ represents the covariance matrix of the observational 
data. If the data points are uncorrelated, then their covariance matrix can be 
expressed as a simple diagonal matrix.

The kernel function of a Gaussian process is a critical component that determines both 
the prior and posterior shapes of the Gaussian process. Each kernel function possesses 
distinct characteristics and can capture various types of patterns. The covariance 
function dictates the smoothness and variability of the predictive outcomes. Among 
the commonly used kernel functions is the squared exponential covariance function 
\cite{williams2006gaussian}, expressed as:
\begin{equation}
  k(x,\widetilde{x})=\sigma_f^2 exp\left [ -\frac{(x-\widetilde{x})^2 }{2\ell ^2}  \right ].
  \label{eq:24}
\end{equation}
Here, $\sigma_f$ and $\ell$ are referred to as hyperparameters. $\sigma_f$ denotes the correlation 
strength, representing the typical variation of the signal in the $f(x)$ direction, while $\ell$ signifies the 
correlation length, indicating the distance along the $x$ direction over which significant changes in the 
function value $f(x)$ occur. They do not directly determine the form of the correlation function but influence 
the fluctuations in function values based on the degree of correlation between the data.
The optimal hyperparameters are obtained from maximizing the logarithm of the marginalized 
likelihood function. The logarithmic form of the marginalized likelihood function is 
expressed as:
\begin{equation}
  \begin{aligned}
  \ln \mathcal{L} = &-\frac{1}{2} (\boldsymbol{y-\mu})^T[K(\boldsymbol{X, X})+C]^{-1}(\boldsymbol{y-\mu})\\
  &-\frac{1}{2}\ln\left |K(\boldsymbol{X,X})+C   \right |-\frac{N}{2} \ln2\pi,
  \label{eq:25} 
  \end{aligned}
\end{equation}
where $N$ is the number of data points.

\begin{figure*}[htbp]
  \centering
  \includegraphics[width=1.0\textwidth]{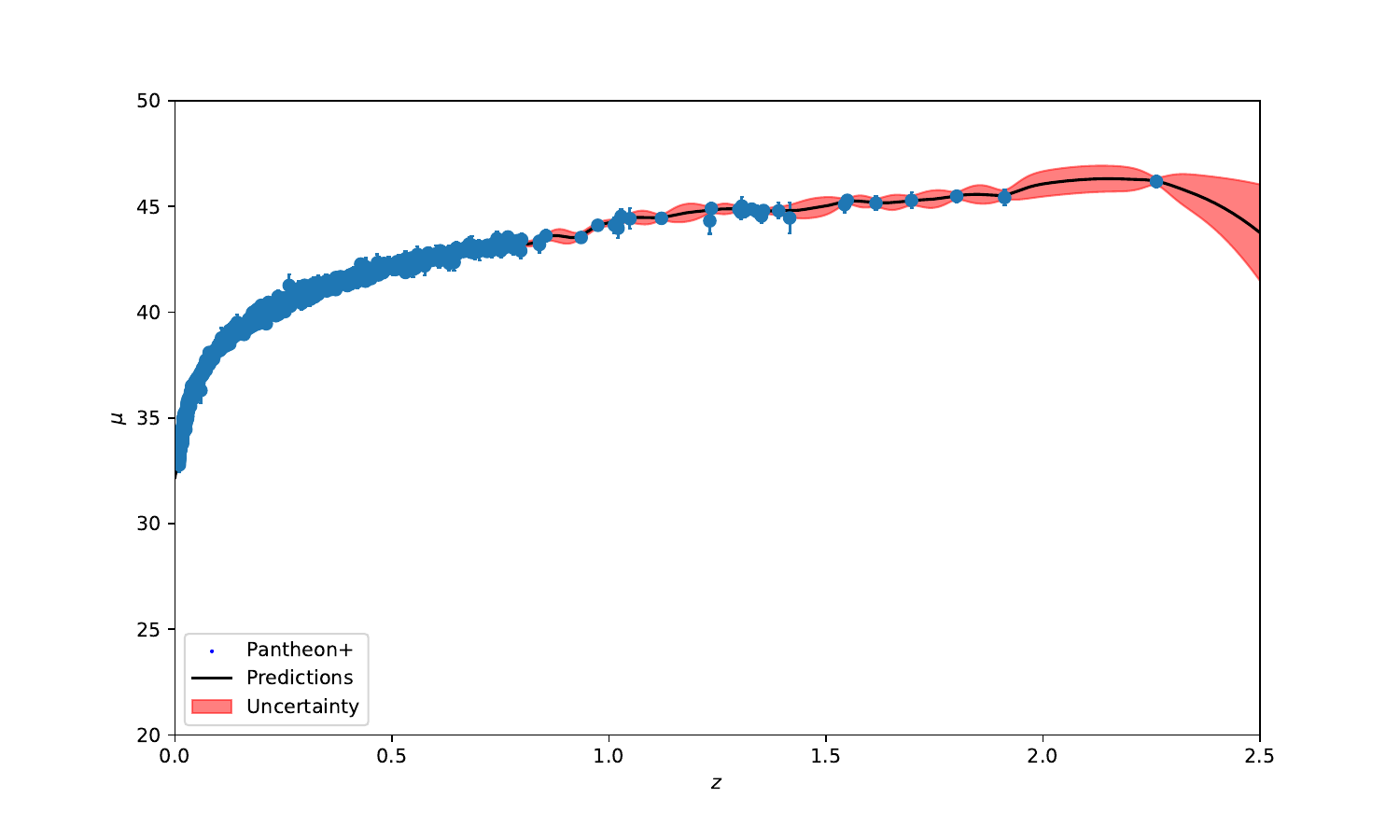}
  \caption{GaPP reconstructed $z-\mu$ plots, where the blue dots are the SNe Ia Pantheon+ data points and 1$\sigma$ 
  errors, the black line is the GaPP reconstructed type Ia supernovae data, and the red region is partly the 
  1$\sigma$ errors of the reconstructed data.}
  \label{fig:2}
\end{figure*}

Next, we proceed with the Gaussian Process reconstruction of the selected low-redshift probes. 
Initially, we employ GaPP to reconstruct the relationship between redshift $z$ and Hubble 
parameter $H(z)$ using the Hubble dataset. Given the interdependence among certain 
observations within this dataset, careful handling of both the covariance matrix and 
observational data is paramount. Specifically, the Hubble dataset comprises 32 data 
points, where the first 17 data points exhibit diagonal covariance matrices, while 
the subsequent 15 data points demonstrate correlated measurements. The correlated 
covariance matrix can be accessed at \url{https://gitlab.com/mmoresco/CCcovariance/}. 
For our analysis, we utilize the complete covariance matrix.


By examining the effect of different kernel functions on the reconstruction of the Hubble dataset, 
we find that as the differentiability of the kernel increases (large $\nu$ in the Matérn class), 
we expect the reconstruction results to be smoother. This is due to stronger correlations and smaller 
errors or confidence intervals \cite{2021elucidatingcosmologicalmodeldependence}. 
Ultimately, we chose Matérn ($\nu = 3/2$) for 
reconstructing $H(z)$. This choice not only provides similar mean function results to those obtained 
with other kernel functions but also ensures that the reconstruction results adhere to the 
maximum error budget.

The expression for the Matérn ($\nu = 3/2$) covariance function is:
\begin{equation}
    k(x,\widetilde{x} ) = \sigma_{f}^2exp\left [ -\frac{\sqrt{3}\left | x-\widetilde{x}  \right |  }{\ell }  \right ]\left ( 1+ \frac{\sqrt{3} \left | x-\widetilde{x}  \right | }{\ell } \right )  ,
    \label{eq:3}
\end{equation}

The results of Gaussian process reconstruction $z-H (z)$ are shown in Fig.~\ref{fig:1}.

Therefore, we can know the value of $H(z)$ corresponding to each redshift, and the value of photometric 
distance $d_L$ corresponding to each redshift $z$ can be obtained by numerical integration calculation through 
Eq.~(\ref{eq:1}), and according to the error transfer formula, we can get the error value of photometric distance 
with the expression:
\begin{equation}
  \begin{aligned}
    \sigma_{d_L}^2= &\left \{ \left  [ c\int_{0}^{z}  \frac{1}{H(z')}dz'+c(1+z) \frac{1}{H(z)}  \right ] \sigma_{H(z)}(\frac{\partial H(z)}{\partial z}  )^{-1}\right \}^2\mid _{z=z_i} .
    \label{eq:}
  \end{aligned}
\end{equation}

~\citet{Moresco_2020} analyzed OHD through simulations within the redshift range $0<z<1.5$, ~\citet{10.1093/mnras/staa3926} 
calibrated the Amati relation with OHD for $z<1.43$, ~\citet{li2023testing} set a redshift cutoff at $z = 1.4$ 
when calibrating GRBs with the Hubble parameter. They considered the Amati relation to be correct, using this 
empirical luminosity relation to constrain cosmological parameters. In contrast, the purpose of this paper is to 
detect the Amati relation and its possible correction in the low redshift range, so we use the entire sample of 
low redshift (i.e., the Hubble dataset with a redshift $z<1.965$) to calibrate low redshift gamma-ray bursts. The 
value of the luminosity distance $\log(d_L)$and its error $\sigma_{\log({d_L})}$(its redshift range is 
$0.0331<z<1.95$) for 48 data points in the sample of Gamma-Ray Burst A118 sample.

Next, we reconstruct the relation between the redshift $z$ of type Ia supernovae and the distance modulus $\mu$ by 
Gaussian process. Here, we choose the double squared exponential covariance function, to ensure that the reconstructed 
function does not exhibit weird oscillations~\cite{mu2023cosmography}. The expression for the covariance function is as follows:
\begin{equation}
  k(x,\tilde{x} )=\sigma_{f1}^2\exp\left [ -\frac{(x-\tilde{x})^2}{2\ell_1 ^2}  \right ] +\sigma_{f2}^2\exp\left [ -\frac{(x-\tilde{x})^2}{2\ell_2 ^2}  \right ].
  \label{eq:5}
\end{equation}

The Gaussian process reconstruction of $z-\mu$ results is shown in Fig.~\ref{fig:2}. Then, we can know the value of 
the distance modulus $\mu$ corresponding to each redshift and its error.

For a given cosmological model, the expression of the distance modulus for predicting redshift is:
\begin{equation}
  \mu_{th} = {5\log[{d_L(z,\theta )\over {\mathrm{Mpc}}}] + 25}.
  \label{eq:6}
\end{equation}

Then the value of $\log(d_L)$ can be found with the expression:
\begin{equation}
  \log(d_L) = (\mu- 25)/5.
  \label{eq:7}
\end{equation}
According to the error transfer, the error is: $\sigma_{\log(d_L)}={1\over5} \sigma_{\mu}$.

In Refs.\cite{liang2022calibrating,mu2023cosmography}, it was noted that the reconstruction function exhibits significant 
uncertainty within the range of $1.4<z<2.3$, potentially impacting the comprehensive analysis of data comparison. 
Consequently, redshifts $z<1.4$ were utilized to calibrate the empirical luminosity relation of GRBs. To maintain 
consistency with the maximum redshift of the Hubble dataset, sample points with SNe Ia redshifts less than 1.95 
were employed. According to cosmological principles, identical redshifts correspond to identical luminosity 
distances, allowing us to determine the value of the luminosity distance $\log(d_L)$ and its error $\sigma_{\log({d_L})}$
for the 48 data points in the Gamma-Ray Burst A118 sample.

\section{Selecting the best Amati relation to expand the Hubble diagram of GRBs}
\label{sec:3}

The standard Amati relation~\cite{amati2002intrinsic,amati2006p,amati2008measuring} is based on the empirical 
luminosity relation between the peak energy of the still frame ($E_{p}$) and the isotropic energy ($E_{iso}$).
Here, $E_p=E_p^{obs}(1+z)$, where $z$ represents the redshift of the GRB, and $E_p^{obs}$ is a 
parameter obtained from the detector database. The expression for the isotropic equivalent energy $E_{iso}$ 
emitted by the GRB can be defined as follows:
\begin{equation}
    E_{iso}=\frac{4\pi {d_L}^2S_{bolo} }{1+z} ,
    \label{eq:8}
\end{equation}
where, ${d_L}$ denotes the luminosity distance, while ${S_{bolo}}$ represents the bolometric flux of GRBs. 

In the previous discussion, we reconstructed the relation between the Hubble parameter $H(z)$ and SNe Ia 
redshift and distance modulus respectively by Gaussian process, and calculated the luminosity distance value and 
its error of low redshift gamma-ray bursts. 
Subsequently, we computed the luminosity distance values and their associated uncertainties for 
low-redshift gamma-ray bursts (GRBs). With reference to Eq.~(\ref{eq:8}), we are able to deduce 
the corresponding $E_{iso}$ values for each redshift $z$.

When fitting the calibration relation, the Amati relation can be expressed as a linear regression relation:
\begin{equation}
    y_i = a + bx_i ,
    \label{eq:9}
\end{equation}
where “$x_i$” is a mathematical function that is related to the peak energy $E_{p,i}$, 
while “$y_i$” is a mathematical function that is associated with the radiative flux $S_{bolo}$
and distance.
\begin{subequations}
	\begin{align}
	&x_i = \log{{E_{p,i}}\over 300\mathrm{keV}},\\
	&y_i = \log{{E_{iso}}\over {\mathrm{erg}}} = \log{{4 \pi d_L^2 S_{bolo,i}} \over 1+z} + 2\log{{d_L}}.
	\end{align}
	\label{eq:10}
\end{subequations}

The likelihood function for gamma-ray bursts is:
\begin{equation}
    \mathcal{L} \propto  {\prod_{i}^{N}}\frac{1}{\sqrt{2\pi \sigma_{tot,i}^2} } \times \exp \left ( -\frac{(y_i-a-bx_i)^2}{2\sigma_{tot,i}^2}  \right )  ,
    \label{eq:11}
  \end{equation}
where $\sigma_{tot,i}^2= \sigma_{ext}^2+\sigma_{y_i}^2+b^2{\sigma_{x_i}}^2$, based on the principle of error propagation,
\begin{subequations}
	\begin{align}
		&\sigma_{x,i} = {{\sigma_{E_{p,i}}}\over \ln10E_{p,i}}, \\
		&\sigma_{y,i} = {{\sigma_{S_{bolo,i}}}\over \ln10S_{bolo,i}}.   
	\end{align}
	\label{eq:12}
\end{subequations}

One of the extended forms of the Amati relation, proposed by ~\citet{wang2017evolutions}, takes into 
account the significant statistical evolution observed in the empirical luminosity relation. They suggest that 
the correlation coefficient may evolve with redshift. To address this issue, inspired by studies on the evolution 
of fitting parameters in light curves of supernovae observations and the intrinsic luminosity evolution of gamma-ray 
bursts (GRBs), they introduce two additional redshift-related terms to represent the redshift evolution of the 
luminosity empirical relation. To avoid the evolution of the correlation coefficient of the Amati relation at high 
redshifts, two moderate formulas are chosen to extend the Amati relation, with their respective expressions being:
\begin{equation}
    a\rightarrow {A = a+{\alpha z\over {1+z}}}, b\rightarrow {B = b+{\beta  z\over {1+z}}},
    \label{eq:13}
\end{equation}
so, the expanded form of the Amati relation becomes as follows:
\begin{equation}
    y_i = (a+{\alpha z\over {1+z}}) + (b+{\beta  z\over {1+z}})x_i,
    \label{eq:14}
\end{equation}
here, $x_i$ and $y_i$ adhere to the expressions outlined in Eq.~(\ref{eq:10}) of the standard Amati relation. 
The likelihood function remains identical to the standard form. However, the parameters $a$ and $b$ in the likelihood 
function are aligned with Eq.~(\ref{eq:13}).

Another extension of the Amati relation is proposed by ~\citet{demianski2021prospects}, who introduces 
redshift-evolution terms into the standard Amati relation, assuming $g_{iso}(z)=(1+z)^{k_{iso}}$ and
 $g_p(z)=(1+z)^{k_p}$.
Then, the de-evolved isotropic equivalent energy $E_{iso}'={E_{iso} \over {g_{iso}(z)}}$, and 
$E'_{p,i}={E_{p,i} \over {g_p(z)}}$. The extended Amati relation becomes:
\begin{equation}
    y_i = a + bx_i + (k_{iso}-bk_p)\log(1+z).
    \label{eq:15}
\end{equation}

Of course, we can simplify the above equation by introducing a single average coefficient $c$, namely:
\begin{equation}
    y_i = a + bx_i + c\log(1+z),
    \label{eq:16}
\end{equation}
where $x_i$ and $y_i$ also adhere to the expressions outlined in Eq.~(\ref{eq:10}) of the standard Amati 
relation. However, in this case, the likelihood function for gamma-ray bursts is modified to:
\begin{equation}
    \begin{aligned}
    \mathcal{L} &\propto  {\prod_{i}^{N} }\frac{1}{\sqrt{2\pi \sigma_{tot,i}^2} } \\
                &\times \exp \left ( -\frac{(y_i-a-bx_i-c\log(1+z_i))^2}{2\sigma_{tot,i}^2}  \right ). \\
    \label{eq:17}
    \end{aligned}
\end{equation}
The $\sigma_{tot,i}^2$ remains consistent with the expression in the standard Amati relation.

In Section \ref{sec:2}, we obtained the distance modulus corresponding to the low redshift of gamma-ray bursts 
by reconstructing the low-redshift calibration stars through a Gaussian process. Next, we utilize the open-source 
software package Cobaya~\cite{torrado2021cobaya} as the sampling tool for Markov Chain Monte Carlo (MCMC) sampling. 
MCMC is a powerful statistical technique employed for sampling from complex probability 
distributions. This method utilizes the Metropolis-Hastings algorithm and, aided by a uniform prior probability 
distribution, generates a chain of sample points to sample from the parameter space.

Firstly, we sampled different bins of gamma-ray bursts with MCMC to obtain Amati and its possible modified 
correlation coefficients. Then, MCMC was used to globally fit the three relations. The results are shown in
 Table~\ref{Tab:1}.

Based on Table~\ref{Tab:1}, it is observed that 
for the standard Amati relation, the constraint results of its free parameters in different bins are consistent 
within the 1$\sigma$ confidence level, indicating no significant evidence for the redshift evolution of the Amati relation. 
In contrast, for the two extended Amati relations, the tension between the parameter constraints from low-redshift 
and high-redshift data is alleviated due to the inclusion of additional parameters to describe redshift dependence, 
as compared to the standard Amati relation. However, the introduction of extra free parameters may introduce considerable 
uncertainty. As shown in Table~\ref{Tab:1}, the errors in the coefficients of the two extended Amati relations are larger, 
which could weaken their constraining power. 

Furthermore, through the calibration of the empirical luminosity relations of 
GRBs with $H(z)$ and the Pantheon+ samples, the central values of the redshift evolution coefficients in  
the two extended Amati relations encompass 0 within the 1$\sigma$ range across different bins. Therefore, to further compare these 
three empirical luminosity relations, a global fitting was performed over the entire redshift range, and the Akaike Information 
Criterion (AIC) and Bayesian Information Criterion (BIC) were calculated for comparison.

The Akaike Information Criterion (AIC)~\cite{akaike1974new,akaike1981likelihood} is a standard for evaluating 
the complexity of statistical models and measuring the goodness of fit of statistical models. It assesses the 
relative merits of models by considering the maximum likelihood estimate of the model and the number of parameters 
in the model. In AIC, penalties for model complexity increase with the number of parameters, thereby avoiding the 
selection of overly fitted models. In general, it is expressed as:
\begin{equation}
   \mathrm{AIC}  = 2k-2\ln(\mathcal{L}),
  \label{eq:18}
\end{equation}
where $k$ is the number of fitting parameters, and $\mathcal{L}$ is the maximum value of the likelihood 
function. Specifically, the AIC value approaches negative infinity as the goodness of fit tends to its limit, 
with smaller numerical values representing better ability of the model to explain the data.

The Bayesian Information Criterion (BIC)~\cite{schwarz1978estimating} is another statistical criterion employed 
for model selection. It assesses models by simultaneously considering their goodness of fit and complexity, thereby 
facilitating the selection of the optimal model through the computation of the probability function and the inclusion 
of a penalty term for the number of parameters in the model. This approach helps guard against overfitting 
while providing a balanced method for model selection. Analogous to the Akaike Information Criterion (AIC), 
its expression is:
\begin{equation}
  \mathrm{BIC} = k\ln{N}-2\ln(\mathcal{L}),
  \label{eq:19}
\end{equation}
in this equation, $k$ and $\mathcal{L}$ correspond to their definitions within the expression for the 
Akaike Information Criterion (AIC), while $N$ represents the number of observed data points utilized 
in the analysis.

We computed the AIC and BIC values for each of these three relations separately. Typically, model 
comparisons are made based on the differences in AIC (or BIC), denoted as $\Delta$AIC (or $\Delta$BIC). 
Table~\ref{Tab:1} summarizes the $\Delta$AIC and $\Delta$BIC values, representing the differences in AIC and 
BIC relative to the reference model (the standard Amati relation).

For the values of $\Delta$AIC and $\Delta$BIC, if $\Delta$AIC($\Delta$BIC) < 1, it indicates that there is little 
difference between the two models according to statistical model selection criteria. 
If 1 < $\Delta$AIC($\Delta$BIC) < 5, the models are not very close, and there may be a preference 
for the model with smaller $\Delta$AIC($\Delta$BIC). If $\Delta$AIC($\Delta$BIC) > 5, the difference 
between the models is very significant. By computing $\Delta$AIC and $\Delta$BIC, as shown in Table~\ref{Tab:1}, 
we can observe that the original standard Amati relation is favored over the two extended Amati 
relations. Therefore, in the subsequent calculations, we will continue to use the standard Amati 
relation. When calculating the luminosity distance at high redshifts, we will apply the correlation 
coefficients obtained from the 48 GRB samples along with their associated errors for the Amati relation.

According to Eq.~(\ref{eq:10}b) and the error propagation formula, we obtain:
\begin{equation}
  (\log{d_L})_i^{data}=\frac{1}{2} (a+b\frac{\log{E_{p,i}}}{300{\mathrm{keV}}}-\log{\frac{4{\pi}S_{bolo,i}}{1+z} } ),
  \label{eq:20}
\end{equation}
\begin{equation}
  \begin{aligned}
    [\sigma(\log{d_L})_i^{data}]^2&=\frac{1}{2} [\sigma_{ext}^2+\sigma_a^2+(\sigma_b\frac{\log{E_{p,i}}}{300{\mathrm{keV}}})^2\\
    &+ (b\frac{\sigma_{E_{p,i}}}{\ln10{E_{p,i}}})^2+(\frac{\sigma_{S_{bolo,i}}}{\ln10S_{bolo,i}} )^2].\\
    \label{eq:21}
  \end{aligned}
\end{equation}
By incorporating the coefficients and uncertainties of the Amati relation obtained via MCMC for 48 GRBs calibrated 
with two low-redshift calibration objects, we can derive the logarithmic luminosity distances and their errors for 
70 data points at high redshifts ($1.9685 < z < 8.200$).

To extend the Hubble diagram of gamma-ray bursts to high redshifts, we calculate the distance modulus at high 
redshifts based on Eq.~(\ref{eq:6}). 
According to error propagation, it can be derived that:
\begin{equation}
  \sigma_{\mu}=5\sigma_{\log{d_L}}.
  \label{eq:22}
\end{equation}

\begin{figure*}[ht]
  \centering
  \includegraphics[width=0.9\textwidth]{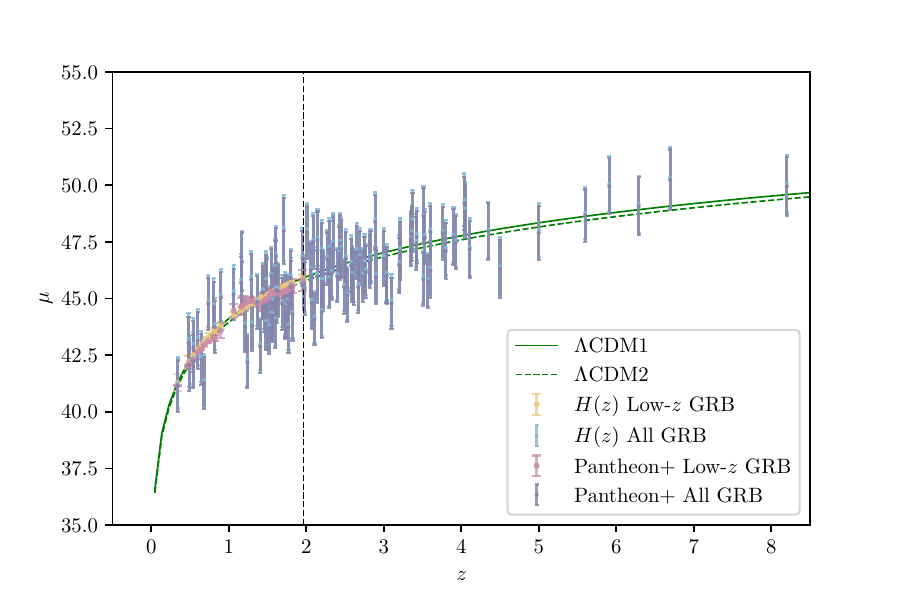}
  \label{fig:3}
  \caption{Hubble diagram of the A118 sample of gamma-ray bursts, where the yellow dots are the distance moduli 
  of the sample points of the 48 low-redshift gamma-ray bursts obtained from the reconstructed $H(z)$ data, the 
  blue dots are the distance moduli of the sample of GRBs A118 computed by extending the correlation coefficients 
  of the low-redshift Amati relation obtained by calibrating the $H(z)$ to high redshifts. The pink dots are the 
  distance moduli of the sample of GRBs A118 obtained from the reconstructed SNe Ia are the distance moduli 
  of the 48 sample points of gamma-ray bursts at low redshift, and the purple dots are the distance moduli of 
  the GRBs A118 sample computed by expanding the correlation coefficients of the low-redshift Amati relation 
  obtained by calibrating the SN to high redshift. The solid green curve is the $H(z)$ standard distance modulus with
  $H_0=69.43$ $\mathrm{km}$ $\mathrm{s^{-1}}$ $\mathrm{Mpc^{-1}}$, $\Omega_m=0.309$. 
  The dashed green curve is the SNe Ia standard distance modulus with $H_0=73.6$ $\mathrm{km}$ $\mathrm{s^{-1}}$ $\mathrm{Mpc^{-1}}$, $\Omega_m=0.334$. 
  The black dotted line denotes z = 1.965.} 
\end{figure*}

Through calculations, we can determine the distance modulus corresponding to the redshift of the gamma-ray 
burst sample A118. The Hubble diagram of gamma-ray bursts is illustrated in Fig.~\ref{fig:3}.

\section{Results and discussion}
\label{sec:4}
In this paper, based on the concept of the distance ladder, we examine the Amati relation and its possible 
corrections for the higher-quality gamma-ray burst A118 sample, selecting a more reliable 
empirical photometric relation, which in turn leads to a model-independent calibration of 
the gamma-ray bursts, expanding the Hubble diagrams to higher redshift.

We reconstructed the Hubble parameter $H(z)$ and Pantheon+ samples separately using GaPP, obtaining the distance 
modulus for each redshift. Based on the principles of cosmology, celestial objects with the same redshift possess 
the same luminosity distance. Therefore, we were able to derive the corresponding luminosity distances for the 48
low-redshift gamma-ray bursts. Considering that the Amati relation may evolve with redshift, Wang et al. and Demianski et al. 
extended this empirical luminosity relation differently. Therefore, we binned the samples at low redshift of GRBs. 
Through MCMC sampling, we obtained the correlation coefficients for the empirical luminosity relations in 
different bins. From Table~\ref{Tab:1}, we observe that the correlation coefficients of the Amati relation are 
consistent within the 1$\sigma$ confidence level across different bins, indicating no significant evidence for redshift 
evolution of the standard Amati relation (at the 68$\%$ confidence level). For the other two extended Amati relations, 
the redshift evolution coefficients include 0 within the 1$\sigma$ range. Subsequently, to further 
assess their behavior across the entire redshift range, we performed a global fit for these three relations 
using MCMC sampling and calculated their respective AIC and BIC. By comparing these values with those of the 
standard Amati relation, we found that the results of $\Delta$AIC and $\Delta$BIC suggest that,  for different 
low-redshift calibrated gamma-ray bursts A118 sample, the standard Amati relation can already adequately fit the data 
compared to potential modifications of the Amati relation.

\renewcommand{\arraystretch}{2.0}
\begin{landscape}
  \begin{table}[htbp]  
  \centering
    \caption{Calibration of the gamma-ray burst (GRB) standard Amati relation, Wang’s Amati relation, and Demianski’s Amati 
    relation using the Hubble data set and Pantheon+ samples, including correlation coefficients and 68$\%$ confidence level 
    errors from MCMC analysis across different bins.
    It is worth noting that the $\Delta$ AIC and $\Delta$ BIC in the table represent the 
    differences in AIC and BIC between the standard Amati relation and the other two Amati relations for GRBs.}
    \centering
    \begin{tabular}{cccccccccc}
      \hline 
             &               &  $a$                       &  $b$                       &  $c$                  &  $\alpha$            &  $\beta$                &  $\sigma_{ext}$                 &  $\bigtriangleup$AIC    &  $\bigtriangleup$BIC \\  \hline
             {Hubble data set}\\ \hline
             & 24Low-$z$ GRBs  &  $52.90\pm 0.13$         &  $1.10\pm 0.20$           &  \                     &     \               &     \                    &  $0.493^{+0.058}_{-0.092}$      &           \             &         \             \\  
  Amati      & 24High-$z$ GRBs &  $53.03\pm 0.14$         &  $1.24\pm 0.22$           &  \                     &     \               &     \                    &  $0.408^{+0.065}_{-0.082}$      &           \             &         \              \\  
             & 48All GRBs      &  $52.970\pm 0.071$       &  $1.18\pm 0.14$           &  \                     &     \               &     \                    &  $0.445^{+0.043}_{-0.057}$      &           -             &         -               \\  \hline
  Wang's     & 24Low-$z$ GRBs  &  $52.53\pm 0.54$         &  $0.8^{+1.0}_{-1.3}$      &  \                     & $0.8\pm 1.1$        & $0.6^{+2.6}_{-1.9}$      &  $0.505^{+0.063}_{-0.096}$      &           \             &         \                \\  
  Amati      & 24High-$z$ GRBs &  $52.5^{+1.8}_{-1.4}$    &  $1.1^{+1.4}_{-2.6}$      &  \                     & $0.8^{+2.1}_{-2.9}$ & $0.3^{+3.7}_{-2.8}$      &  $0.407^{+0.055}_{-0.087}$      &           \             &         \                 \\  
             & 48All GRBs      &  $52.49\pm 0.37$         &  $0.62\pm 0.93$           &  \                     & $0.88\pm 0.67$      & $ 0.9\pm 1.7$            &  $0.448^{+0.043}_{-0.058}$      &      13.072482          &     16.814884              \\  \hline
  Demianski's& 24Low-$z$ GRBs  &  $52.61\pm 0.37$         &  $1.06\pm 0.20$           &  $1.0\pm 1.3$          &     \               &     \                    &  $0.497^{+0.058}_{-0.093}$      &           \             &         \                   \\
  Amati      & 24High-$z$ GRBs &  $52.7^{+1.3}_{-0.79}$   &  $1.24\pm 0.22$           &  $0.8^{+1.7}_{-3.3}$   &     \               &     \                    &  $0.411^{+0.052}_{-0.089}$      &           \             &         \                    \\
             & 48All GRBs      &  $52.60\pm 0.28$         &  $1.12\pm 0.14$           &  $1.08\pm 0.78$        &     \               &     \                    &  $0.436^{+0.041}_{-0.055}$      &      3.654998           &     5.526199                  \\  \hline
             {Pantheon+ samples}\\ \hline
          
           & 24Low-z GRBs  &  $52.86\pm 0.10$        &  $1.13\pm 0.20$           &  \                     &     \                  &     \                    &  $0.499^{+0.061}_{-0.094}$      &           \             &         \             \\  
Amati      & 24High-z GRBs &  $52.96\pm 0.13$        &  $1.27\pm 0.23$           &  \                     &     \                  &     \                    &  $0.408^{+0.054}_{-0.087}$      &           \             &         \              \\  
           & 48All GRBs    & $52.916\pm 0.071$       &  $1.20\pm 0.14$           &  \                     &     \                  &     \                    &  $0.446^{+0.040}_{-0.054}$      &           -             &         -               \\  \hline
Wang's     & 24Low-z GRBs  &  $52.39\pm 0.53$        &  $0.8\pm 1.2$             &  \                     & $0.99\pm 1.1$          & $0.5\pm 2.2$             &  $0.509^{+0.063}_{-0.098}$      &           \             &         \                \\  
Amati      & 24High-z GRBs &  $52.5^{+2.0}_{-1.2}$   &  $1.2\pm 1.8$             &  \                     & $0.7^{+2.1}_{-3.0}$    & $0.2\pm 2.9$             &  $0.406^{+0.053}_{-0.087}$      &           \             &         \                 \\  
           & 48All GRBs    & $52.42^{+0.33}_{-0.39}$ &  $0.61\pm 0.92$           &  \                     & $0.90^{+0.70}_{-0.61}$ & $0.9\pm 1.6$             &  $0.441^{+0.042}_{-0.056}$      &      13.12239           &     16.865792              \\  \hline
Demianski's& 24Low-z GRBs  &  $52.51\pm 0.37$        &  $1.06\pm 0.21$           &  $1.3\pm 1.3$          &     \                  &     \                    &  $0.506^{+0.061}_{-0.094}$      &           \             &         \                   \\
Amati      & 24High-z GRBs &  $52.6^{+1.2}_{-0.86}$  &  $1.26\pm 0.23$           &  $0.8^{+1.7}_{-3.3}$   &     \                  &     \                    &  $0.407^{+0.054}_{-0.088}$      &           \             &         \                    \\
           & 48All GRBs    & $552.55\pm 0.27$        &  $1.13\pm 0.15$           &  $1.05\pm 0.75$        &     \                  &     \                    &  $0.438^{+0.040}_{-0.057}$      &      3.873152           &     5.744353                   \\  \hline 
              
  \label{Tab:1}
    \end{tabular}
    \end{table}            
  \end{landscape}

Therefore, we extend the correlation coefficient of the low redshift of the standard Amati relation directly 
to the high redshift, and further calculate the distance modulus of the gamma-ray burst at the high redshift, 
so we extend the Hubble diagram to the higher redshift. 

Certainly, due to the limited number of data points and the presence of different empirical luminosity relations 
for various samples of GRBs, there is considerable controversy regarding whether the correlation coefficients evolve 
with redshift, as noted in Refs.\cite{8150632,Tsutsui_2009}. Therefore, in order to calibrate gamma-ray bursts as 
standard candles for cosmological probes, further investigation into whether their empirical luminosity relations 
evolve with redshift is warranted. We look forward to additional observational data and improved calibration methods 
to facilitate a more definitive assessment.

\begin{acknowledgements}
  This work is supported in part by National Natural Science Foundation of China under Grant No. 12075042 and No. 11675032.
\end{acknowledgements}

\bibliography{paper}

\end{document}